 \documentclass[final,3p,times]{elsarticle}

\usepackage{graphicx}
\usepackage{subfigure}
\usepackage[utf8]{inputenc}
\usepackage{indentfirst}
\usepackage{hyperref}
\hypersetup{colorlinks, linkcolor=black, citecolor=black, urlcolor=black}
\usepackage{breakurl}

\journal{arXiv}

\begin{document}
\begin{frontmatter}



\title{Two-colour generation in a chirped seeded Free-Electron Laser}

\author[Elettra,CEA,UNG]{B. Mahieu}\ead{benoit.mahieu@cea.fr}
\author[Elettra]{E. Allaria}
\author[Elettra]{D. Castronovo}
\author[Elettra]{M. B. Danailov}
\author[Elettra]{A. Demidovich}
\author[UNG,Elettra]{G. De Ninno}
\author[Elettra]{S. Di Mitri}
\author[Elettra]{W. M. Fawley}
\author[Elettra]{E. Ferrari}
\author[Elettra]{L. Fröhlich}
\author[UNG]{D. Gauthier}
\author[Elettra,Frascati]{L. Giannessi}
\author[Elettra]{N. Mahne}
\author[Elettra]{G. Penco}
\author[Elettra]{L. Raimondi}
\author[Elettra]{S. Spampinati}
\author[Elettra]{C. Spezzani}
\author[Elettra,UT]{C. Svetina}
\author[Elettra]{M . Trov\`{o}}
\author[Elettra,CNR]{M . Zangrando}

\address[Elettra]{Elettra - Sincrotrone Trieste, S.S.14 - km 163.5 in AREA Science Park, 34149 Basovizza, Italy}
\address[CEA]{Service des Photons Atomes et Mol\'{e}cules, Commissariat \`{a}  l'Energie Atomique, Centre d'Etudes de Saclay,
B\^{a}timent 522, 91191 Gif-sur-Yvette, France}
\address[UNG]{Laboratory of Quantum Optics, University of Nova Gorica, Vipavska 11c, 5270 Ajdov\v{s}\v{c}ina, Slovenia}
\address[Frascati]{Theory Group - ENEA C.R. Frascati, Via E. Fermi 45, 00044 Frascati, Italy}
\address[UT]{University of Trieste, Graduate School of Nanotechnology, Piazzale Europa 1, 34127 Trieste, Italy}
\address[CNR]{IOM-CNR, S.S.14 - km 163.5 in AREA Science Park, 34149 Basovizza, Italy}

\begin{abstract}
We present the experimental demonstration of a method for generating two spectrally and temporally separated pulses by an externally seeded, single-pass free-electron laser operating in the extreme-ultraviolet spectral range. Our results, collected on the FERMI@Elettra facility and confirmed by numerical simulations, demonstrate the possibility of controlling both the spectral and temporal features of the generated pulses. 
A free-electron laser operated in this mode becomes a suitable light source for jitter-free, two-colour pump-probe experiments.

\end{abstract}

\end{frontmatter}


\section*{Introduction}

\label{intro}

Chirped pulse modulation is a widely used technique, with relevant applications in several fields of research, ranging from digital communication \cite{DC} to radar detection \cite{RD}, and from optics \cite{OP} to laser physics. In a broadband laser, frequency chirping is employed to stretch a short pulse prior to amplification. This stretching and the associated lower intensity mitigates the problem of phase distortion in the amplification medium. After pulse amplification, the pulse is recompressed to recover a short duration and, hence, high power \cite{CPA}. A similar method has been proposed \cite{YuPRE, NIMPoletto, NIMFeng} to enhance the power generated by a single-pass seeded free-electron laser (FEL) working in the so-called coherent harmonic generation (CHG) regime \cite{YuPRA}.  
After amplification and removal of the chirp, one in principle can obtain a high power, nearly Fourier-transform-limited pulse.  


In this paper, we characterize the performance of a seeded FEL where frequency chirping on the seed laser is instead exploited to generate two independent pulses spectrally and temporally separated in an externally controllable fashion. As previously shown in \cite{PRLDP}, this opens the possibility of using a chirped-seeded FEL as a self-standing source for pump-probe experiments in the vacuum ultraviolet and X-ray spectral domain.      

The paper is divided into the following parts: in Section \ref{formation}
we introduce the principle upon which a CHG FEL is based, with specific reference to the case of the FERMI@Elettra FEL \cite{NaturePhot}. We then qualitatively describe the formation of the double spectral peak, when the seed laser carries a linear frequency chirp. In Section \ref{description}, we give a description of the effects due to both a laser frequency chirp and an electron beam energy chirp. Then in Section \ref{experiment}, we present the results obtained on the FERMI@Elettra FEL and, in Section \ref{simulations}, compare them to numerical simulations. Finally, we summarize our conclusions and provide some perspectives for future experimental studies.

\section{Formation of the spectral double peak}

\label{formation}

\begin{figure}[!h]
\centering
\includegraphics[trim = 0mm 60mm 0mm 50mm, clip=true, width=0.9\textwidth]{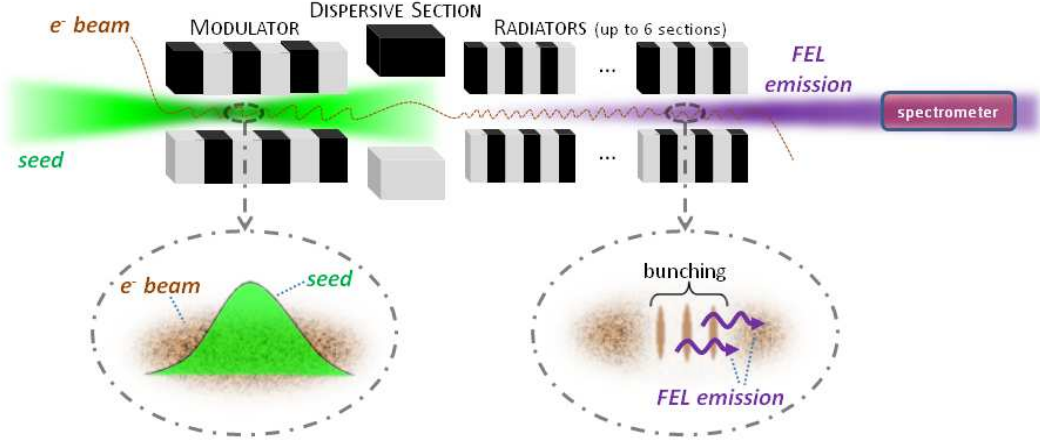}
\caption{Schematic layout of the CHG configuration of FERMI@Elettra (see text for details on the FEL process). The output light is separated downstream the FEL line by a grazing-incidence mirror. The fraction corresponding to zeroth order of diffraction ($\ge 95\%$ of the total) is sent to experimental user beamlines, whereas the much weaker, first order component is sent to an online diagnostic spectrometer where it strikes a YAG:Ce crystal, whose visible fluorescence is then measured by a charge-coupled device (CCD) \cite{PADRES1,PADRES2}. The spectrometer resolution is about $1.4\,$meV per CCD pixel.}
\label{Figure_FERMI}
\end{figure}

A CHG FEL such as FERMI@Elettra typically employs a layout similar to that shown in Fig. \ref{Figure_FERMI}. A relativistic electron beam first propagates through the periodic field generated by an undulator, called the modulator, simultaneously interacting with a powerful coherent light source (e.g., a low-order harmonic of a Ti:Sapphire laser). 
This interaction coherently modulates the electron energy which then leads to a longitudinally-regular charge distribution after the electrons pass through a chromatically-dispersive magnet. Such a spatial modulation, called bunching, has nonzero spectral components at both the seed wavelength and at its higher harmonics.
After the dispersive section, electron beam enters one or more additional undulators (called the radiators), whose normalized magnetic deflection parameters $K$ are tuned for fundamental FEL resonance to occur at an integral harmonic of the seed wavelength: 
\begin{equation}
\label{resonance}
\lambda_n=\frac{\lambda_u}{2 n \gamma^2}\, (1+\frac{K^2}{2}).
\end{equation}
Here $\lambda_u$ is the undulator period, $n$ the harmonic order and $\gamma$ is the relativistic Lorentz factor characterizing the mean energy of the electron beam.

The resultant FEL radiation, which is highly temporally coherent due to the periodic bunching, can undergo quadratic (low-gain regime) or exponential (high-gain regime) amplification. For a sufficiently long radiator, the FEL process will reach saturation.
The main parameters of the FERMI@Elettra configuration relevant for our experimental studies are listed in Table \ref{Table_FEL1_parameters} and its caption.
One should note that on FERMI@Elettra the seed pulses are much shorter than the electron beam.
This is a critical point both for the spectral quality of the FEL emission and the spectro-temporal pulse-splitting phenomenon. 

\begin{table}[!h]
\centering 
\small
\begin{tabular}{ccc}
\hline\hline
Electron beam & Mean energy & $\approx 1\,$GeV \\
& Peak current & $\approx 300\,$A \\
& Duration (FWHM) & $1.5\,$ps \\
\hline
Seed & Central wavelength & $261\,$nm\\
& Temporal profile & Gaussian \\
& Duration (FWHM) & $ \approx 200\,$fs \\
& Energy & $10 - 150\,\mu$J \\
& Transverse size (RMS) & $\approx 350 \,\mu$m \\
\hline
Modulator & Period & $100\,$mm\\
& Number of periods & $32$\\
\hline 
Dispersive section & $R_{56} $ & $20 - 50\,\mu$m \\ 
\hline
Radiators & Number & 2 \\
& Harmonic order $n$ & 6 \\
& Period & $55\,$mm \\
& Number of periods & $44$\\
\hline\hline
\end{tabular}
\caption{Parameters of FERMI@Elettra relevant for the experiments reported in this paper.
The seed transverse size is defined as the standard deviation of the transverse intensity distribution measured at the modulator entrance. The $R_{56}$ parameter is the momentum compaction factor characterizing the strength of the dispersive section \cite{YuPRA}. The undulators were set so as to generate circularly polarized FEL radiation, with a typical energy/pulse of  $20\,\mu$J.}
\label{Table_FEL1_parameters}
\end {table}

\begin{figure}[!h]
\centering
\includegraphics[trim = 0mm 2mm 0mm 0mm, clip=true, width=0.7\textwidth]{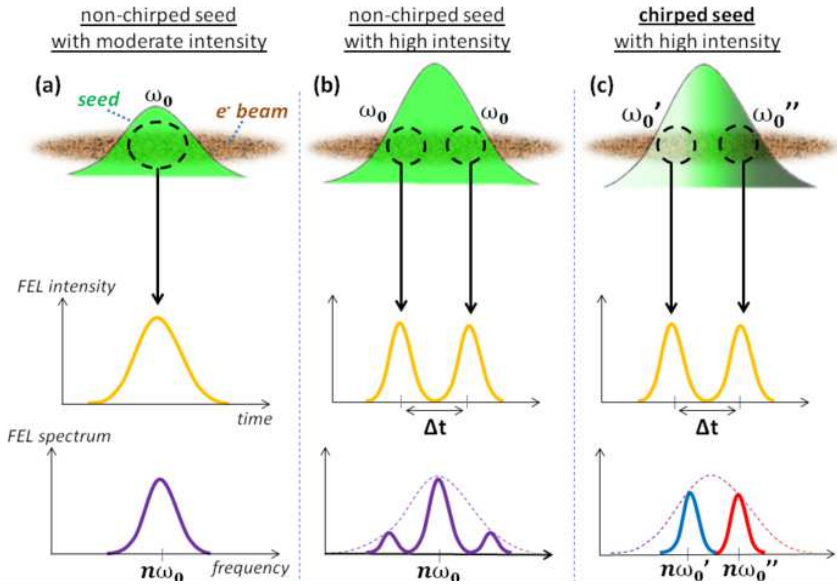}
\caption{Sketch of the seed-electrons interaction and resulting FEL (temporal and spectral) outputs for different seed configurations: no chirp and moderate seed intensity (a), no chirp and high seed intensity (b), chirped seed with high intensity (c). In (a), the optimum seed-electrons interaction occurs around the center of the seed pulse so that the FEL output mimics the shape of the seed. In (b), the FEL pulse temporally splits in two because the seed power is too high in the middle of the pulse: a beating between the two sub-pulses involves a frequency modulation but the spectrum remains centered on a single peak at the harmonic $n$ of the constant seed frequency $\omega_0$. In (c), the chirp of the seed combined with the temporal pulse splitting leads to the creation of two separated spectral peaks corresponding to the harmonics of the frequencies $\omega_0'$ and $\omega_0''$ at the respective position of each sub-pulse.}
\label{Figure_double_peak_principle}
\end{figure}

As shown in Fig.~\ref{Figure_double_peak_principle}a, under conditions of moderate seed power and a temporally flat (i.e, both in energy and current) electron beam, the spectro-temporal properties of CHG mimic those of the seed \cite{PRLSR, NaturePhot}: starting with a Gaussian seed pulse with a given duration and wavelength $\lambda$, one ends up with an output Gaussian CHG pulse at wavelength $\lambda_n$ with slightly shorter duration\cite{Stupakov} .        
However, this statement (which, in the temporal domain, is mainly a prediction of numerical simulations, due to the lack of temporal diagnostics in the vacuum-ultraviolet spectral range), is true only under certain limits and depends on the regime of operation of the FEL.
Indeed, when increasing the seed power, the temporal edges of the seed pulse are also able to produce strong bunching at entrance to the radiator so that a larger part of the electron beam emits intense FEL radiation, giving a longer FEL pulse. 
For very high seed powers, the electron-beam energy modulation at the peak of the seed (i.e., the central position for a Gaussian) becomes too strong for the particular $R_{56}$ value characterizing the strength of the dispersive section. Then there is an overbunching effect and the resultant bunching in the radiator is strongly depressed. Thus, the FEL emission from this central position falls. 
On the other hand, there are two temporal positions with lower seed power, symmetric with respect to the peak of the seed,  that match the optimum bunching condition for a given $R_{56}$. 
Accordingly, the FEL signal grows and reaches the maximum intensity in these regions. This scenario is represented in Fig. \ref{Figure_double_peak_principle}b. For a sufficiently strong seed intensity, the FEL power at the central position almost vanishes and two successive pulses appear, as previously described in \cite{Labat}.
This phenomenon becomes particularly interesting if the seed is chirped i.e., if its frequency depends on the longitudinal position. As shown in Fig. \ref{Figure_double_peak_principle}c, in this situation the split temporal pulses have also different frequencies. In other words, two independent pulses with two different frequencies are created \cite{PRLDP}. 

\begin{figure}[!h]
\centering
\resizebox{0.8\textwidth}{!}{\includegraphics{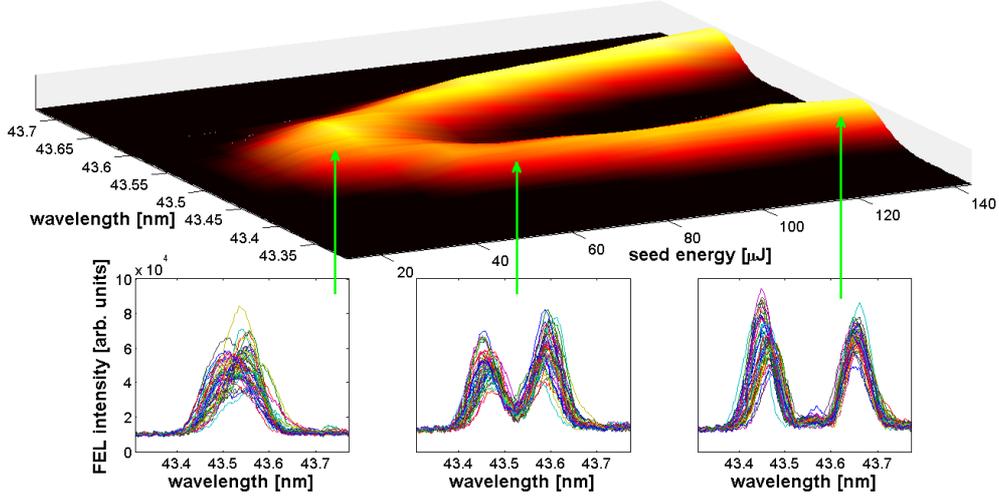}}
\caption{Experimental characterization of the spectral separation between FEL peaks, as a function of seed energy per pulse. The insets show the spectrum of fifty consecutive shots, integrated at the positions indicated by the arrows on the projected figure (higher part).}
\label{Figure_fork_seed_experiment}
\end{figure}

Two-colour pulse generation by a seeded FEL was observed for the first time \cite{talkAllaria} early in 2011 during the commissioning of FERMI@Elettra. A typical measurement of the FEL spectrum as a function of the seed power is shown in Fig. \ref{Figure_fork_seed_experiment}. By increasing the seed energy, the portion of electrons getting over-bunched increases, leading to a larger separation of the spectral peaks. As shown in the figure insets, the two-colour spectra  are very stable on a shot-to-shot basis with very good repeatability. We obtained very similar pulse-splitting results by increasing the strength of the dispersive section while keeping the seed power constant (this is illustrated later in Fig. \ref{Figure_R56_experiment}).

The spectro-temporal properties of the FEL emission are also affected by the quality of the electron beam. In particular, effects due to an energy chirp (i.e., the mean electron energy varies with longitudinal position) can couple with those from the frequency chirp carried by the seed laser. This interplay may have an impact on the process of formation of the two peaks, as well as on their spectral and temporal separations. 
In order to better explain the process of production of the double peak, we will now provide a description of the seed laser and electron beam chirps, with particular reference to the case of FERMI@Elettra.

\section{Description of seed and electron beam chirps}
\label{description}

\subsection*{Seed pulse: frequency chirp}

The seed pulse used at FERMI@Elettra is the third harmonic of a Ti:Sapphire laser; both the temporal and spectral profiles are well fit by a Gaussian dependence \cite{seedFERMI}. The presence of a linear frequency chirp can be modelled by a quadratic term in the phase of the laser electric field:
\begin{equation}
\label{Efield_t}
E(t) = A(t) e^{i(\omega_0 t + \Gamma_i t^2)}.
\end{equation}
Here A(t) stands for the temporal envelope while the exponential term carries the phase of the oscillations, including the rapidly-oscillating term at the central angular frequency $\omega_0$ and the quadratic term $\Gamma_i t^2$. The parameter $\Gamma_i$ is given by:
\begin{equation}
\label{gamma}
\Gamma_i = \pm\frac{1}{4\sigma_t^2}\sqrt{4\sigma_t^2 \sigma_{\omega}^2-1},
\end{equation}
$\sigma_t$ and $\sigma_{\omega}$  being the standard deviations of the temporal intensity profile and the spectrum, respectively. The sign of $\Gamma_i$ corresponds to the sign of the induced frequency dispersion. In the setup adopted at FERMI@Elettra, the frequency dispersion of the seed pulse may be controlled in three different (but eventually combined) ways: (1) by propagation of the seed through  different materials; (2) by passage through a dedicated set of gratings; and (3) by increasing the seed intensity, which causes self-phase modulation \cite{Diels}. It is worth noting that in the presence of self-phase modulation, the chirp cannot be considered as being linear at or near the edges of the pulse. Moreover, the self-modulation process is associated with a spectral broadening. This broadening has been measured and taken into account in the present study. We also note that, for the experiments reported in this paper, the Gaussian shape of the pulse is preserved over the full range of seed energies. 

The instantaneous frequency of the pulse is the derivative of the temporal phase (see Eq. \ref{Efield_t}):
\begin{equation}
\label{frequency}
\omega(t) = \omega_0 + 2\Gamma_i t.
\end{equation}
Positive $\Gamma_i$’s define a positive linear chirp; i.e., smaller frequencies (longer wavelengths) arrive first. On the contrary, negative chirps arise for negative $\Gamma_i$ values. 
In both cases, two parts of the pulse separated by a duration $\Delta t$ have  different frequencies, whose separation is, according to Eq. \ref{frequency}:
\begin{equation}
\label{dw}
\Delta\omega(\Delta t) = 2\Gamma_i\Delta t.
\end{equation}
For a CHG FEL that operates in the pulse splitting regime, with the radiators tuned at the harmonic $n$ and the two sub-pulses separated by $\Delta t$, we thus expect the frequency separation between the two spectral peaks to be:
\begin{equation}
\label{dwn}
\Delta\omega _n = n\Delta\omega
\end{equation}
For the sake of convenience, in the reminder of this paper, 
we will speak in terms of wavelength $\lambda$, instead of angular frequency $\omega$ 
and will therefore consider $\Delta\lambda _n$ as the spectral separation of the two peaks of the FEL emission.

\begin{figure}[!h]
    \centering
    \includegraphics[width=0.5\textwidth]{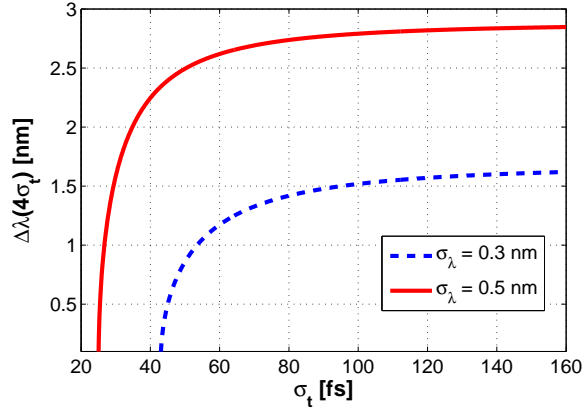}
    \caption{Evolution of $\Delta\lambda(4\sigma_t)$ as a function of $\sigma_t$ for the following bandwidths of the seed carrying a linear positive chirp: $\sigma_{\lambda}=0.3\,$nm (dashed line) and $\sigma_{\lambda}=0.5\,$nm (full line). These values correspond to limiting cases of accessible Gaussian spectra at FERMI@Elettra. For a negative chirp, one gets the same values of $\Delta\lambda(4\sigma_t)$, but with opposite sign. The $\sigma_t$ corresponding to $\Delta\lambda(4\sigma_t) = 0\,$nm is the (theoretical) duration of a Fourier-transform limited pulse.}
    \label{Figure_seed_data_without_alphabeta}
\end{figure}
For a given spectral bandwidth, the presence of chirp in the seed pulse will increase its temporal duration.
At two positions separated by $4\sigma_t$ and symmetric with respect to the center of the pulse, the electron bunch experiences $13.5\%$ of the maximum intensity of a Gaussian profile seed. For our seed laser parameters, efficient seeding seems difficult below $10\%$ of the peak intensity, i.e., on the very far edges of the seed pulse. 
Therefore, we may consider $4\sigma_t$ as the maximum practically achievable $\Delta t$ for a given chirp.
Figure \ref{Figure_seed_data_without_alphabeta} displays the evolution of the spectral separation $\Delta \lambda$ between parts of the seed located at a distance $4\sigma_t$, as a function of $\sigma_t$. Two different values of the bandwidth have been considered, corresponding to the two limiting cases at FERMI@Elettra: for a seed energy per pulse below $10\,\mu$J, the spectral bandwidth is close to $\sigma_{\lambda}=0.3\,$nm whereas, for higher energies and thus intensities, nonlinear effects can broaden the spectrum up to $\sigma_{\lambda}=0.5\,$nm. For increasing values of $\sigma_t$ (i.e., for an increasing amount of chirp), $\Delta \lambda$ reaches an asymptotic value. 
Examining Fig.~\ref{Figure_seed_data_without_alphabeta}, one therefore sees for the case of FERMI@Elettra $\Delta\lambda$ is limited to about $3\,$nm. This information is important because it sets an effective limit to the spectral separation $\Delta\lambda _n$ of the two pulses generated by the FEL.
\\

\subsection*{Electron beam: energy chirp}

As is now well known for CHG FEL's in which the electron beam acts as an intermediary in carrying the coherent seed signal, temporal variations such as chirps in the energy profile may seriously impact the FEL output emission 
(see, e.g., \cite{Biedron_Milton_Freund, Shaftan, Fawley_Penn, LutmanGreen, Lutman}). 
A linear electron beam energy chirp, whose slope will be noted $\chi_{1}$, couples to the total chromatic dispersion of the dispersive section and downstream undulators resulting in a simple
shift of the central wavelength of FEL emission \cite{Shaftan}. 
In principle, any corresponding gain reduction due to this shift can be compensated by either retuning the resonant wavelength of
the undulators (see Eq. \ref{resonance}) or by slightly adjusting the seed wavelength. 
However, a quadratic
chirp component $\chi_{2}$  will result in spectral broadening of output emission \cite{Fawley_Penn, Lutman} and, for sufficiently large magnitudes, can also reduce the FEL gain if the broadening becomes a significant fraction of the FEL's gain bandwidth. 
In principle, here too the effect can be compensated by placing an equal magnitude but negative chirp on the seed pulse \cite{Graves}.
In the context of the subject matter of this paper, namely the formation and control of a double spectro-temporal peak,
a quadratic electron beam energy chirp can either enhance or reduce 
the spectral offset between the two peaks and thus must be carefully taken into account.

A simple linear and quadratic energy chirp can be modelled as:
\begin{equation}
\label{energy}
E(t) = E_0 + \chi_1 t + \chi_2 t^2.
\end{equation}
The exact amplitudes of the $\chi_{1}$ and $\chi_{2}$ coefficients at entrance to the FEL depend upon
various propagation effects as the electron beam is transported through the linear accelerator (linac).
First of all, the radio-frequency (RF) accelerating fields \cite{Wangler} of the successive linac sections (generally phased at or very close to crest) induce a negative quadratic chirp. 
Second, in order to increase the electron beam current (and thus enhance the FEL emission), the electron bunch is time-compressed. 
On FERMI@Elettra, this is done by first imposing a positive linear chirp component via phasing of two upstream linac sections off crest, and then by sending the electron beam through a magnetic chicane that converts the linear energy chirp into longitudinal compression. 
However, this procedure generates an additional negative curvature of the temporal energy profile.
In order to linearize the longitudinal electron-beam phase space during the compression (generally necessary to have a relative constant current profile),
an X-band, 4th high-harmonic RF cavity \cite{Xband} positioned just upstream of the chicane is run close to negative crest to provide a positive contribution to the $\chi_{2}$ coefficient and compensate the aforementioned negative curvature. 
Following compression, strong longitudinal wakefields \cite{longwake} in the downstream linac act upon the much higher current electron beam bunch: they induce a chirp that contains both
a linear, negative term and a positive curvature component. 
While the linear $\chi_{1}$ coefficient can be eliminated
by slightly dephasing off crest the last couple RF linac sections (at a price of a small reduction in final beam energy), there is no simple way to reduce the quadratic $\chi_{2}$ coefficient and therefore in general the electron beam enters the FEL with a quadratic energy chirp.

\begin{figure}[!h]
    \centering
    \subfigure
    {
        \includegraphics[trim = 10mm 0mm 15mm 0mm, clip=true, width=0.32\textwidth]{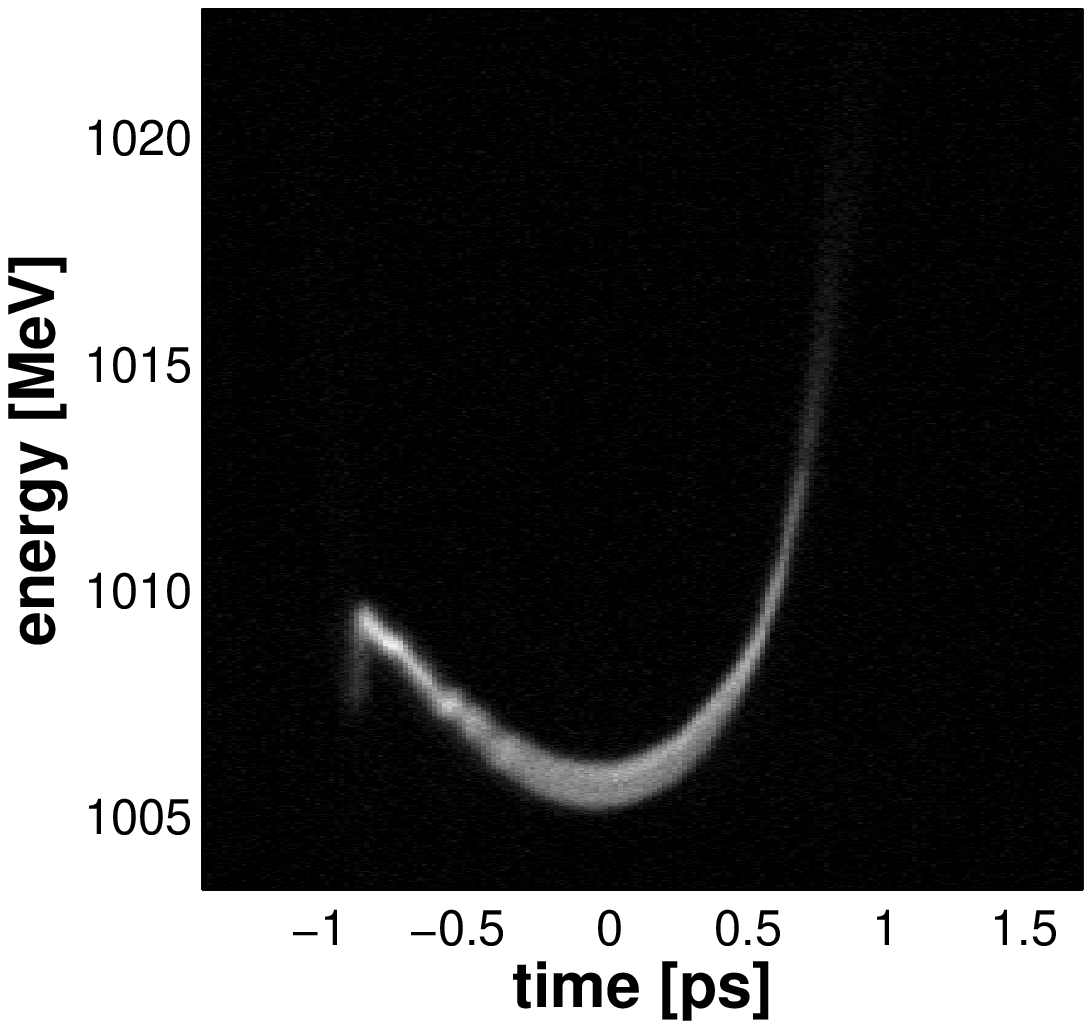}
    }
    \subfigure
    {
        \includegraphics[trim = 15mm 0mm 15mm 0mm, clip=true, width=0.32\textwidth]{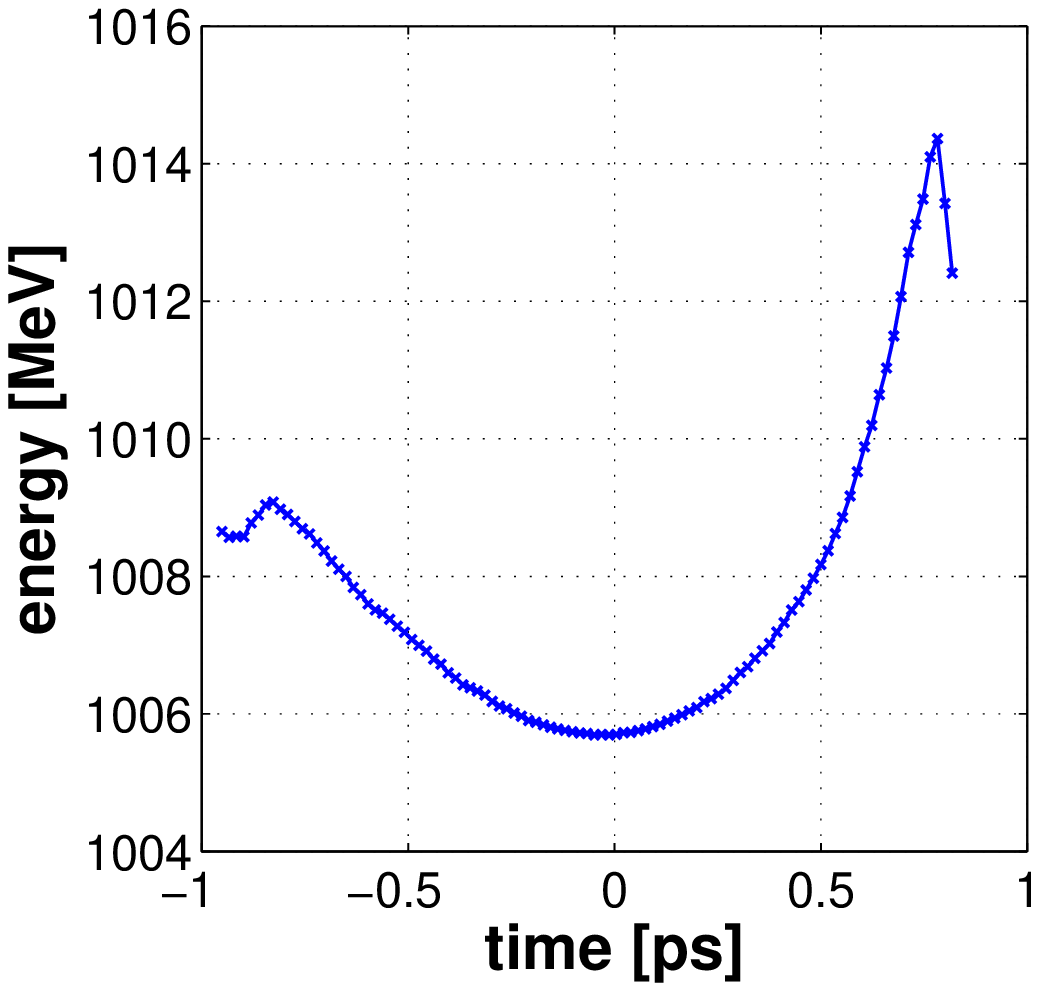}
    }
    \subfigure
    {
        \includegraphics[trim = 15mm 0mm 15mm 0mm, clip=true, width=0.30\textwidth]{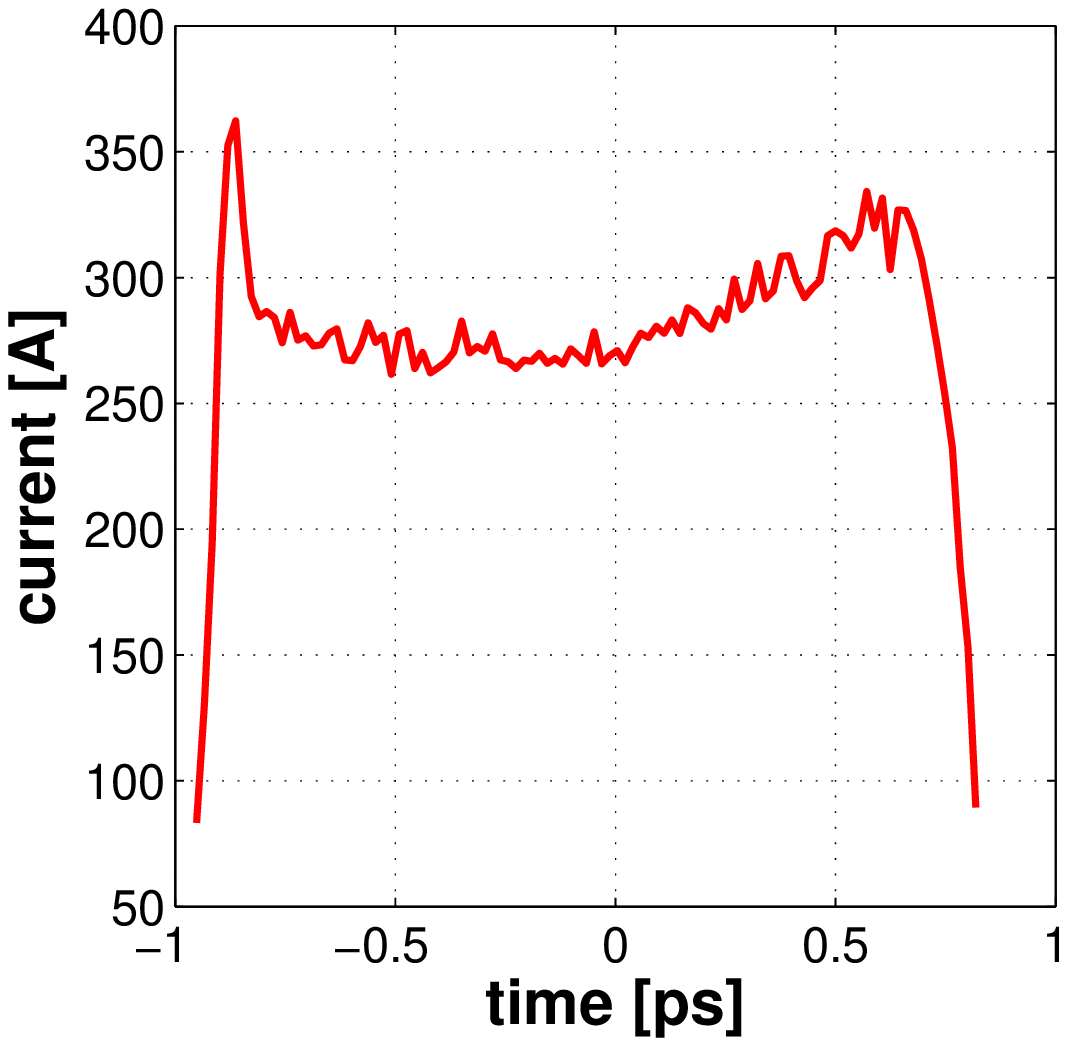}
    }
\caption{Electron beam longitudinal phase space measured at the RF deflector/spectrometer diagnostic end station of the FERMI@Elettra linac. Left: raw CCD spectrometer image; center: energy profile determined from image analysis; right: current profile. Here time \textit{increases} from the beam head to the beam tail.}
\label{Figure_ebeam_measurement}
\end{figure}

At FERMI@Elettra, we characterize the electron beam longitudinal
phase space at the end of the linac via a diagnostic section
composed of an RF deflecting cavity \cite{HERFD} followed by an
energy spectrometer. The cavity stretches the beam in
the vertical direction, inducing a correlation between the temporal
distribution and the vertical displacement. The bending magnet chromatically disperses
the electrons in the horizontal plane, making the information about
the beam energy profile available along that direction. The beam longitudinal phase
space can then be visualized on a fluorescent YAG crystal + CCD camera system placed
downstream, as shown in Fig.~\ref{Figure_ebeam_measurement}. 
This particular $E(t)$ distribution corresponds to the configuration in which
most of the results presented in this paper have been obtained. 
The energy profile is very close to a parabola with a quite small linear component 
(i.e., \hbox{\, $\chi_{1} \approx 1\,$Mev/ps}) 
and a much more significant quadratic component with \hbox{$\chi_{2} \approx 7\,$Mev/ps$^{2}$}. 
In Section~\ref{simulations} we consider the impact of this quadratic component upon double peak formation.

As shown in the right panel of Fig.~\ref{Figure_ebeam_measurement}, the current profile is almost flat in the middle of the bunch, even if a slow ramp towards the tail can be noticed.
Experimentally, there is a typical shot-to-shot jitter of $\approx 50\,$fs of the relative electron beam-seed timing. This jitter in conjunction with current inhomogeneities results in a fluctuation of the FEL power, as the seed ``explores'' different temporal portions of the beam. 
This effect may be relevant for user applications but does not affect the formation of the spectro-temporal double peak. 

We now give a quantitative description of the two-colour emission results at FERMI@Elettra.

\section{Experimental results}
\label{experiment}

\begin{figure}[!h]
    \centering
    \includegraphics[width=0.6\textwidth]{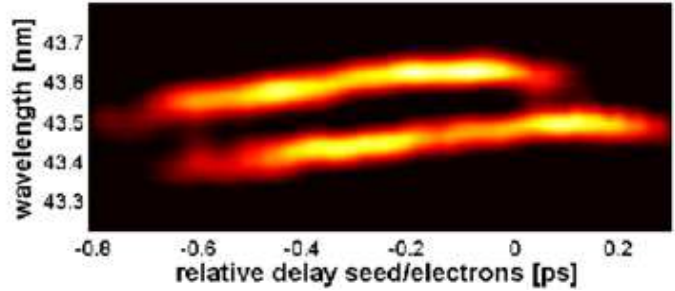}
    \caption{FEL spectrum at the $6^{th}$ harmonic of the seed ($43.5\,$nm), as a function of the relative delay between the seed pulse and the electron beam. At the zero value of the delay, the seed pulse is approximately centered on the bunch of electrons. The dispersive section $R_{56}$ and seed energy were set, respectively, to $23\,\mu$m and $80\,\mu$J; see Table~1 for other parameters.}
    \label{Figure_delay_experiment}
\end{figure}

Figure \ref{Figure_delay_experiment} shows the behaviour of the FEL spectrum at the $6^{th}$ harmonic of the seed (43.5 nm) as one sweeps the relative time delay between the seed and the electron beam. 
To the best of our knowledge, this picture provides the first direct \textit{experimental} evidence of the temporal pulse splitting in an FEL due to local overbunching induced by an intense seed. 
At a delay $ \approx -0.8\,$ps, the seed is superimposed on the electron beam tail, so that the first peak (at larger wavelength, since the frequency chirp is positive) appears. 
Then, at a delay $\approx -0.6\,$ps, the whole seed pulse lies on top of sufficiently high current portions of the electron beam and the second peak (at shorter wavelength) also appears.

As we remarked in the previous section, a linear energy chirp of the electron energy induces a spectral shift of the FEL emission. Hence, when they appear near the tail portion of the electron beam, both peaks are detuned towards shorter wavelengths because of the strong local negative slope of the electron beam chirp (see Fig. \ref{Figure_ebeam_measurement}). 
For smaller relative timing delays (i.e., moving the seed from overlapping the tail toward the head of the bunch), the central wavelengths of the two peaks drift redward due to the increase of the value of the local linear chirp of the electron energy distribution. 
However, the separation between the two spectral peaks ($\Delta\lambda_6 \approx 0.17\,$nm), their relative heights and their respective widths remain practically constant. 
The wavelength drift flattens out about $100\,$fs before the first spectral peak disappears;
this behaviour is mirrored $200\,$fs later in the second peak. This early disappearance might be caused by the small current ramp from the center of the bunch to its tail (see Fig. \ref{Figure_ebeam_measurement}) and/or by degraded transverse properties in this region.

In this experiment, the external seed chirp was created both during the propagation through different optical components such as lenses and windows, and by self-phase modulation due to high intensity. We typically measured a seed bandwidth $\sigma_{\lambda} \approx 0.45\,$nm and a duration $\sigma_t \approx 90\,$fs (determined by cross-correlating the seed pulse with an infrared pulse, extracted from the oscillator of the Ti:Sapphire laser chain, whose FWHM duration $\approx 80\,$fs). This gives $\Gamma_i \approx 6.2 \cdot 10^{-5}\,$fs$^{-2}$, leading to the retrieval of the theoretical separation between the two sub-pulses, defined in Section \ref{description}. From Eqs. \ref{dw} and \ref{dwn}, we obtain:
\begin{equation}
\label{dt}
\Delta t = \frac{\Delta\omega}{2\Gamma_i} = \frac{n\pi c}{\Gamma_i\lambda^2} \cdot \Delta\lambda _n,
\end{equation}
where $c$ is the speed of light. For an harmonic order $n=6$, $\Delta t \approx 230\,$fs (equal to $2.5 \sigma_t$). The accuracy of this result will in principle be affected by the intrinsic chirp induced in the radiation during the FEL process \cite{Wu} and by any quadratic energy chirp on the electron bunch. However, the size of the former is usually quite small relative to other chirps and the contribution of the latter is minor for our conditions, as will be confirmed in Section \ref{simulations}.

\begin{figure}[!h]
    \centering
    \subfigure[]
    {
        \includegraphics[width=0.25\textwidth]{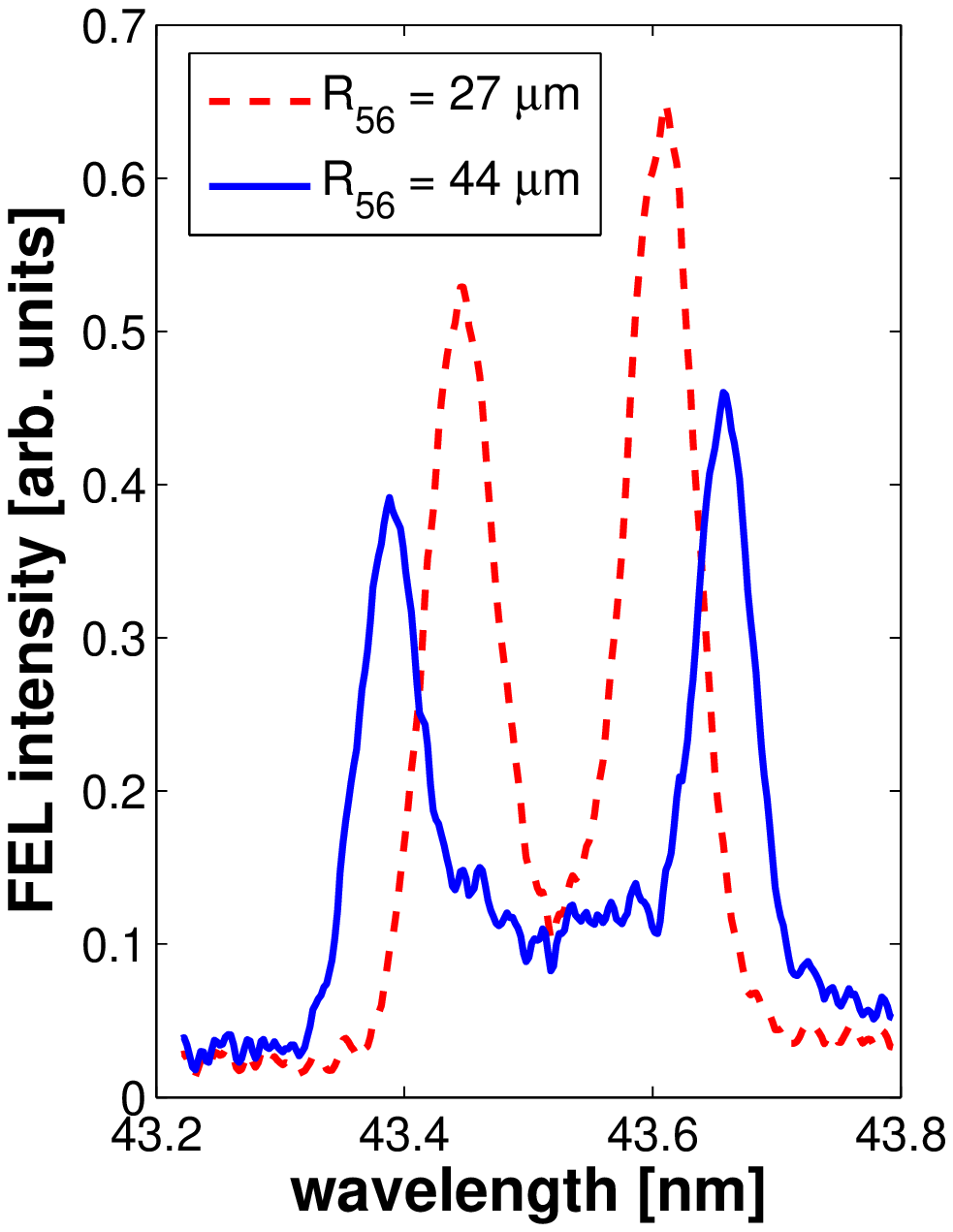}
        \label{Figure_R56_experiment}
    }
    \subfigure[]
    {
        \includegraphics[width=0.25\textwidth]{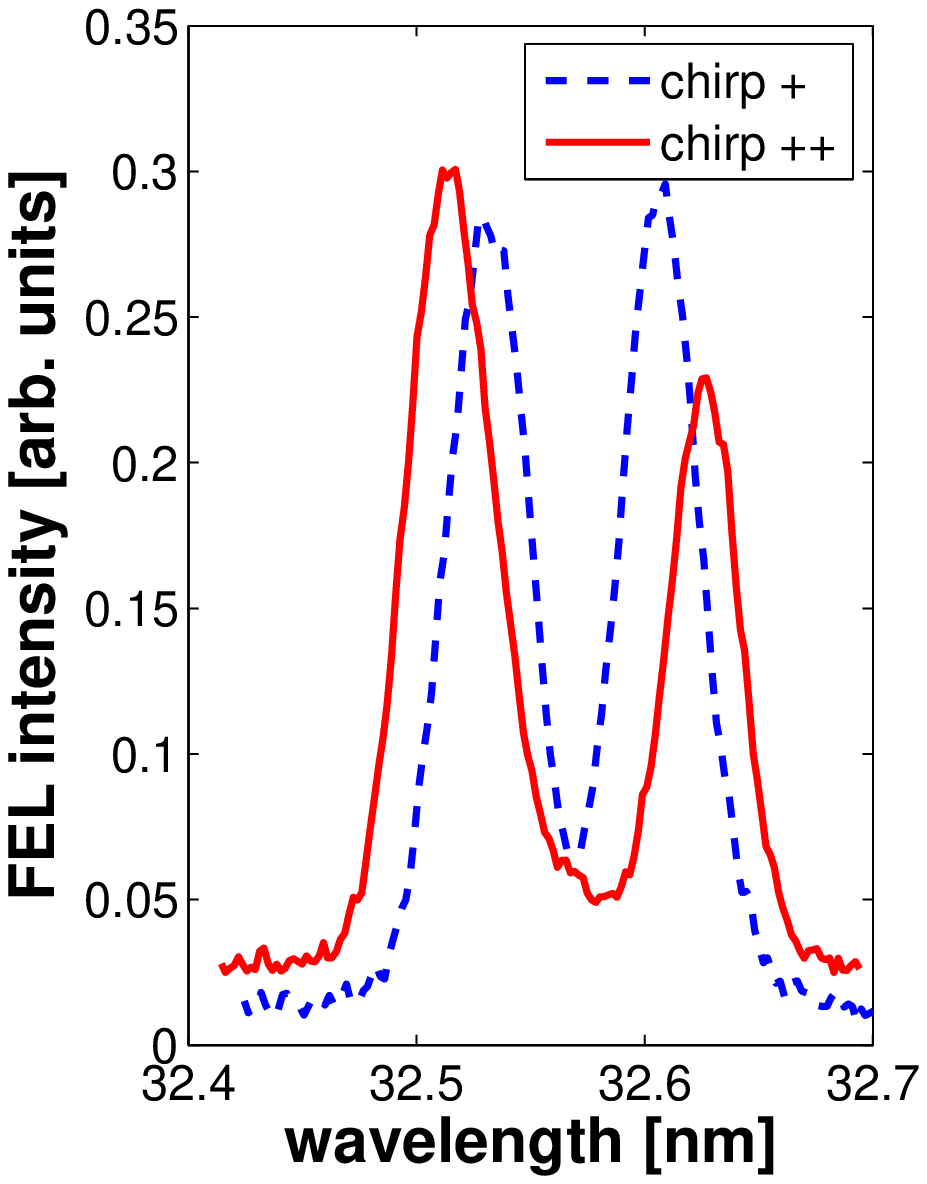}
        \label{Figure_chirp_experiment}
    }
    \caption{Observed double spectral peaks: 
    (a) for two different strengths of the dispersive section and 
    (b) for two different chirp values of the seed (b), corresponding to the standard positive chirp (``chirp +", $\sigma_{\lambda} \approx 0.45\,$nm and $\sigma_t \approx 90\,$fs with a seed energy of $\approx 70\,\mu$J) and a stronger chirp (``chirp ++", $\sigma_{\lambda} \approx 0.5\,$nm and $\sigma_t \approx 160\,$fs with a seed energy of $\approx 75\,\mu$J) induced by adding a piece of 50-mm thickness of $CaF_2$. For (b), the configuration of the machine was different than the one described in Table \ref{Table_FEL1_parameters}: 6 radiators were used, tuned to the $8^{th}$ harmonic of the seed ($32.6\,$nm), $R_{56} = 80~\mu m$ and the electron beam had the following parameters: mean energy $1270\,$MeV, $\chi_1 \approx 8\,$MeV/ps, $\chi_2 \approx 0.3\,$MeV/ps$^2$, FWHM duration $2\,$ps, peak current $\approx 200\,$A.}
    \label{Figure_peaks_games_experiment}
\end{figure}

Up to now, we mainly focused on the effect of the seed power upon double spectro-temporal pulse formation. However, other parameters can influence the production and separation of the peaks.
One is the strength of the dispersive section, as already discussed in \cite{PRLDP} and experimentally demonstrated here in Fig. \ref{Figure_R56_experiment}. Indeed, the dispersive section has an effect very similar to that of the seed power: the higher its strength, the lower the required seed power to produce a double pulse.
Some small differences in the trend may be observed essentially because changing the seed intensity may affect the non-linear optical material dispersion and thus the chirp, while this does not happen when the $R_{56}$ is varied.
Similarly, when employing a strong dispersive section, a moderate seed power can be sufficient to increase the peaks separation. Fig. \ref{Figure_chirp_experiment} shows a different way to modify the spectral separation of the two pulses, i.e., by changing the seed chirp. However, increasing the seed chirp above a given limit does not produce any significant change in the FEL spectrum. This can be expected from the limitation of $\Delta\lambda$ shown in Fig. \ref{Figure_seed_data_without_alphabeta}. 

\begin{figure}[!h]
\centering
\includegraphics[width=0.3\textwidth]{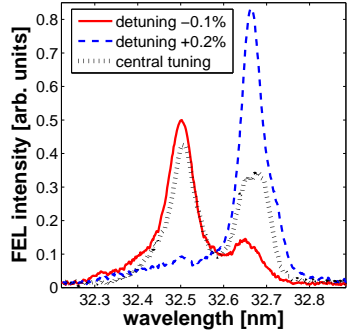}
\caption{Double peak for different tuning conditions of the radiators. The seed was in its standard configuration so that $\sigma_{\lambda} \approx 0.45\,$nm and $\sigma_t \approx 90\,$fs with an energy per pulse of $\approx 70\,\mu$J. The configuration of the rest of the machine was different than the one described in Table \ref{Table_FEL1_parameters}: 6 radiators were used, tuned to the $8^{th}$ harmonic of the seed ($32.6\,$nm), $R_{56} = 60\,\mu$m and the electron beam had the following parameters: mean energy $1175\,$MeV, $\chi_1  \approx 0\,$MeV/ps, $\chi_2 \approx 10\,$MeV/ps$^2$, FWHM duration $0.8\,$ps, peak current $\approx 600\,$A.}
\label{Figure_detuning_experiment}
\end{figure}
    
Also the relative heights of the peaks can be easily controlled, favouring the growth of one over the other, 
by tuning the wavelength corresponding to FEL resonance of the undulators.
The spectrum represented by the dotted line in Fig. \ref{Figure_detuning_experiment} corresponds to a central tuning of the undulators. The data represented by the solid line were taken at slightly negative detuning; while both peaks can be seen, the one at shorter wavelength is significantly stronger. The dashed curve shows a situation where the detuning is positive, for which the peak of the gain curve has moved to longer wavelengths. Here the shorter wavelength pulse has been almost completely suppressed.

\section{Simulations}
\label{simulations}

\begin{figure}[!h]
\centering
    \subfigure[experiment]
    {
        \includegraphics[trim = 20mm 0mm 20mm 0mm, clip=true, width=0.45\textwidth]{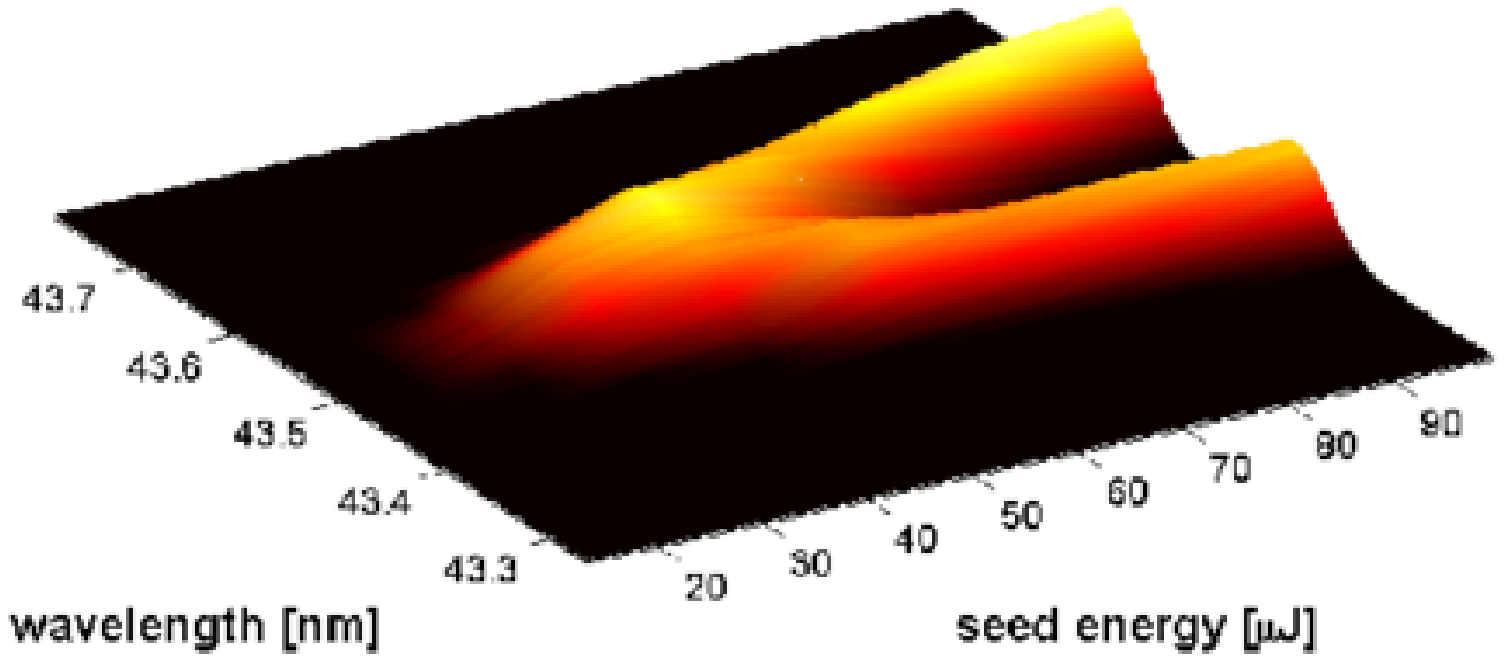}
        \label{Figure_fork_seed_cut_exp_spectrum}
    }
    \subfigure[simulations]
    {
        \includegraphics[trim = 20mm 0mm 20mm 0mm, clip=true, width=0.45\textwidth]{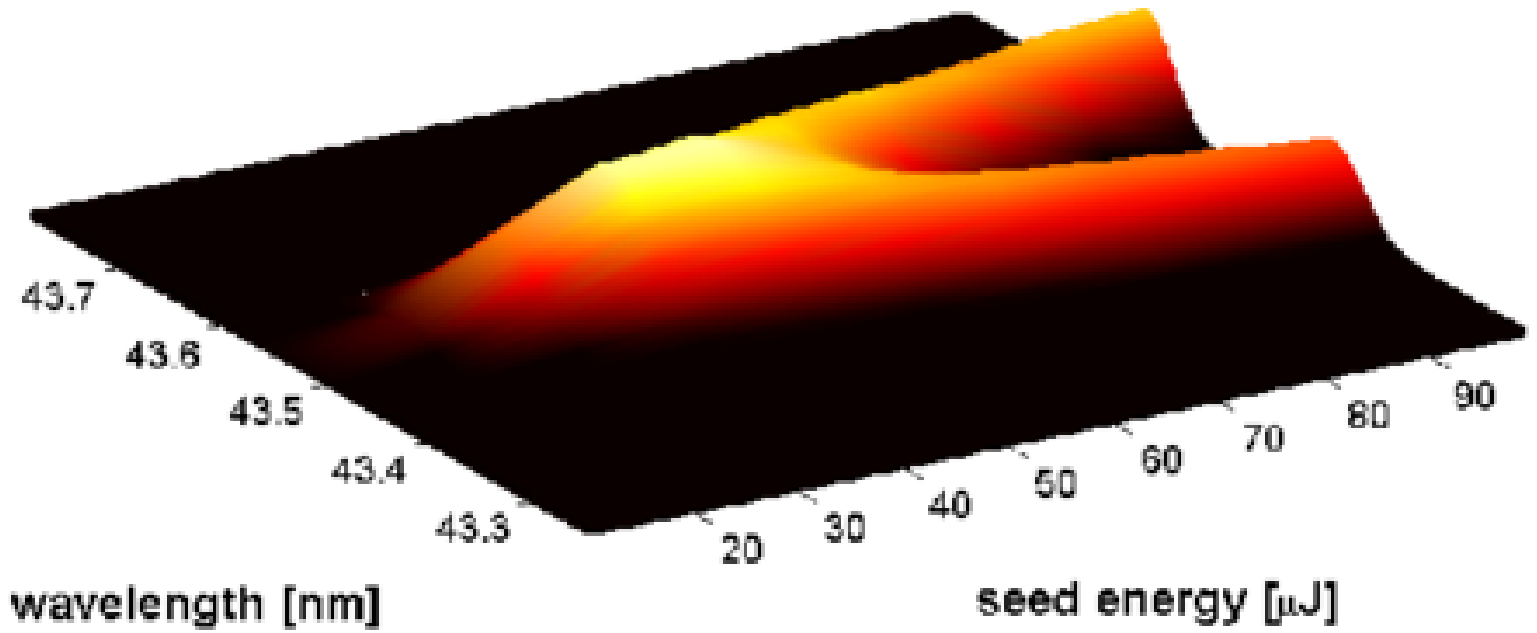}
        \label{Figure_fork_seed_cut_Perseo_spectrum}
    }
\caption{Comparison of the spectral splitting observed experimentally (a) and reproduced by simulations (b) using the code Perseo. For simulations, the following parameters were used for the electron beam: $\chi_{1} = 1\,$MeV/ps, $\chi_{2} = 7\,$MeV/ps$^{2}$, 
normalized transverse emittance $1\,$mm-mrad, relative energy spread $0.01\%$, peak current $300\,$A.
For each value of the seed energy, the bandwidth and the duration of the seed pulse were determined from interpolation of experimental values, growing linearly from $\sigma_{\lambda} = 0.35\,$nm to $\sigma_{\lambda} = 0.45\,$nm and from $\sigma_t = 70\,$fs to $\sigma_t = 90\,$fs as the seed energy increased from $10\,\mu$J to $100\,\mu$J.
For other parameters, please refer to Table \ref{Table_FEL1_parameters}.}
\label{Figure_fork_seed_exp_vs_Perseo}
\end{figure}

The formation of the double peak is well reproduced by simulations \cite{PRLDP} carried out using both 3D \cite{genesis} and 1D numerical codes \cite{perseo}. Figure \ref{Figure_fork_seed_exp_vs_Perseo} shows a comparison of the spectral pulse splitting between the experimental data and 1D Perseo code simulations. As it can be seen, the agreement is very satisfactory. 
Note that in the simulations, the relative power before splitting is slightly higher than the one measured experimentally.

\begin{figure}[!h]
\centering
\includegraphics[width=0.5\textwidth]{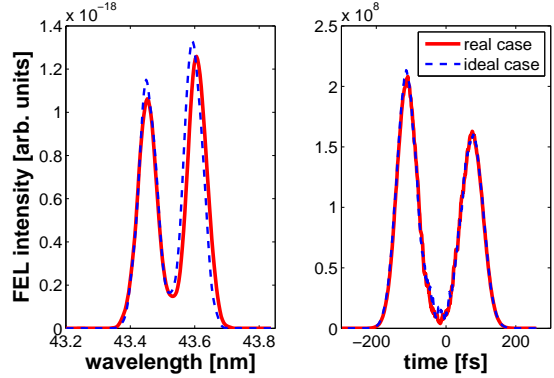}
\caption{Predicted FEL spectral and temporal profiles obtained using the the numerical code Perseo, for the following parameters: seed bandwidth $\sigma_{\lambda} = 0.47\,$nm, duration $\sigma_t = 93\,$fs, energy $70\,\mu$J and central wavelength $261.1\,$nm; $R_{56}= 20\,\mu$m; 
electron beam peak current $300\,$A, normalized transverse emittance $1\,$mm-mrad and relative energy spread $0.01\%$; undulators tuned to the $6^{th}$ harmonic of the seed ($43.5\,$nm). 
The dashed line represents the results obtained using an ideal, constant current, chirpless electron beam. 
The solid line shows simulation results for an electron beam with a current profile corresponding to 
Fig.~\ref{Figure_ebeam_measurement} and energy chirps of $1\,$MeV/ps (linear) 
and $7\,$MeV/ps$^2$ (quadratic).}
\label{Figure_output_Perseo}
\end{figure}

Let us now focus on one single measurement. Figure \ref{Figure_output_Perseo} shows the predicted spectrum and temporal profiles, obtained for simulation parameters that allow a good reproduction of the dashed-line spectrum reported in Fig. \ref{Figure_R56_experiment}. The simulations have been carried out both for an ideal (i.e., flat in current and in energy) electron beam (dashed line), and the ``real'' electron-beam energy distribution shown in Fig. \ref{Figure_ebeam_measurement} (solid line). 
As shown in the left panel of Fig. \ref{Figure_output_Perseo}, the positions of the spectral peaks are similar for the two cases: a slight reduction of the pulse separation on the order of $0.01\,$nm can be observed in the ``real" case. 
Such a small contribution may be neglected in our conditions and, in any case, the double peak formation is not affected. Moreover, if we calculate using Eq. \ref{dt} the theoretical separation between the two sub-pulses in the spectrum represented in Fig. \ref{Figure_R56_experiment}, we find $\Delta t = 198\,$fs. This is close to the value found in the simulation (right panel of Fig. \ref{Figure_output_Perseo}), that is $\Delta t = 186\,$fs. 
Therefore, we claim that the agreement between theoretical, experimental and numerical studies can be considered as very satisfactory. 

For this set of data, the spectral peaks have similar bandwidths $\sigma_{\lambda} \approx 0.028~nm$, and the temporal pulses have similar durations, i.e., $\sigma_t \approx 30~fs$. According to these numbers, the duration of each sub-pulse is $\approx 1.6$ times greater than the Fourier-transform limit for a Gaussian spectrum. The distance from the Fourier-transform limit can be further reduced, e.g. by tuning the chirp of the seed laser. Since the seed and thus the FEL pulses are chirped, one could have expected a significant negative impact on the time-bandwidth product of each sub-pulse. However, this is not the case since each peak is quite narrow, and therefore barely sees the phase curvature of the seed electric field. Moreover, this also indicates that the phase of FEL pulses does not present high-order distortions.

\begin{figure}[!h]
\centering
\includegraphics[width=0.7\textwidth]{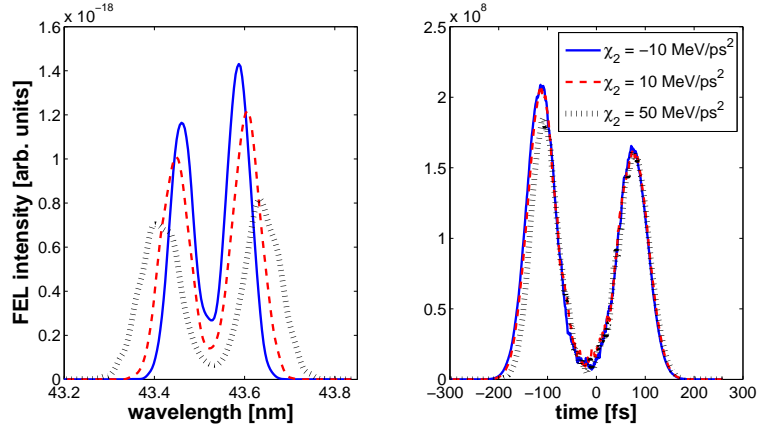}
\caption{Comparison of the outputs of simulations carried out with the numerical code Perseo for different values of the quadratic chirp of the electrons (left: spectrum; right: temporal intensity), for the following parameters: seed bandwidth $\sigma_{\lambda} = 0.47\,$nm, duration $\sigma_t = 93\,$fs, energy $70\,\mu$J and central wavelength $261.1\,$nm; $R_{56}= 20\,\mu$m; undulators tuned to the $6^{th}$ harmonic of the seed ($43.5\,$nm); electron beam peak current $300\,$A, normalized transverse emittance $1\,$mm-mrad, relative energy spread $0.01\%$ and no linear chirp component. 
Solid line: $\chi_2=-10\,$MeV/ps$^2$; dashed line: $\chi_2=10\,$MeV/ps$^2$; 
dotted line: $\chi_2=50\,$Mev/ps$^2$ .}
\label{Figure_quadratic_energy_chirp_Perseo}
\end{figure}

Before concluding, let us consider a situation where the quadratic chirp of the electron beam is no longer negligible. We carried out simulations with same parameters as before for several values of quadratic chirp, but without any linear component (i.e., $\chi_1=0\,$MeV/ps$^2$). 
In Fig. \ref{Figure_quadratic_energy_chirp_Perseo}, we see that the value of $\chi_2$ has almost no effect on the temporal shape of the FEL emission but that, as already stated in Section \ref{description}, the individual bandwidths and separation of the two spectral peaks are directly affected. Consequently, a quadratic energy chirp of the electron beam acts on the spectrum in a very similar way as does a linear chirp of the seed laser, and thus both may be used individually or in conjunction to modulate the spectral shape of the FEL emission.

\section*{Conclusion}
\label{conclusion}

We studied the mechanism leading to the generation of two spectrally- and temporally-separated radiation pulses for a seeded, harmonic upshift free-electron laser 
operating in the extreme ultra-violet spectral region. 
This mechanism, discovered and studied in detail at the FERMI@Elettra FEL facility, 
relies on the use of a high-power chirped seed pulse to produce strong, coherent overbunching in an electron beam; the presence of the chirp is the key element to producing a spectral splitting between the two temporal pulses. 
By changing either the seed laser power or the strength of the dispersive section,
we demonstrated excellent control of both the temporal distance and the spectral separation between the generated pulses. 
We also found that for our particular operation regime,
the onset of the pulse splitting and the control of the pulses properties are only slightly affected 
by temporal variations in electron beam properties such as the current. 
Quadratic energy chirps do affect the spectral bandwidths and separation between the two pulses, but
have negligible effect on their temporal separations.
In the limit that radiation/electron beam slippage effects are small,
the temporal separation between pulses can be directly retrieved from the measurement of the FEL spectrum and from the knowledge of the seed chirp properties.
Such information can be extremely important when exploiting this two-pulse configuration to carry out jitter-free, pump-probe user experiments.

\section*{Acknowledgements}
We gratefully acknowledge the valuable contribution of the FERMI@Elettra commissioning and operations team. We also profited from insightful discussions with F. Bencivenga, C. Callegari, F. Capotondi and M. Labat. This work was funded by the FERMI@Elettra project of Elettra-Sincrotrone Trieste, partially supported by the Ministry of University and Research under grant numbers FIRB-RBAP045JF2 and FIRB-RBAP06AWK3. The work of B. Mahieu, G. De Ninno and D. Gauthier has been partially supported by the CITIUS project, funded by the Program for cross-border cooperation Italy-Slovenia 2007-2013.

%

\end{document}